\documentclass[fleqn,usenatbib]{mnras}

\usepackage{newtxtext,newtxmath}

\usepackage[T1]{fontenc}
\usepackage{ae,aecompl}
\usepackage{datatool}

\usepackage{graphicx}	
\usepackage{amsmath}	
\newcommand\textlcsc[1]{\textsc{\MakeLowercase{#1}}}

\title[Detection of emission from dark matter subhaloes]{Baryonic effects on the detectability of annihilation radiation from dark matter subhaloes around the Milky  Way}

\author[R. J. J. Grand et al.]{\parbox[t]{\textwidth}{
Robert J. J. Grand\thanks{E-mail: grand@mpa-garching.mpg.de}, Simon D. M. White}
\\ 
Max-Planck-Institut f\"{u}r Astrophysik, Karl-Schwarzschild-Str. 1, 85748 Garching, Germany\\
}

\date{Accepted XXX. Received YYY; in original form ZZZ}

\pubyear{2020}

\begin{document}
\label{firstpage}
\pagerange{\pageref{firstpage}--\pageref{lastpage}}
\maketitle

\begin{abstract}
We use six, high-resolution $\Lambda$CDM simulations of galaxy formation to study how emission from dark matter annihilation is affected by baryonic processes. These simulations produce isolated, disc-dominated galaxies with structure, stellar populations, and stellar and halo masses comparable to those of the Milky Way. They resolve dark matter structures with mass above $\sim 10^6$ $\rm M_{\odot}$ and are each available in both full-physics and dark-matter-only versions. In the full-physics case, formation of the stellar galaxy enhances annihilation radiation from the dominant smooth component of the galactic halo by a factor of three, and its central concentration increases substantially. In contrast, subhalo fluxes are {\it reduced} by almost an order of magnitude, partly because of changes in internal structure, partly because of increased tidal effects; they drop relative to the flux from the smooth halo by 1.5 orders of magnitude. The expected flux from the brightest Milky Way subhalo is four orders of magnitude below that from the smooth halo, making it very unlikely that any subhalo will be detected before robust detection of the inner Galaxy. We use recent simulations of halo structure across the full $\Lambda$CDM mass range to extrapolate to the  smallest (Earth-mass) subhaloes, concluding, in contrast to earlier work, that the total annihilation flux from Milky Way subhaloes will be less than that from the smooth halo, as viewed both from the Sun and by a distant observer. Fermi-LAT may marginally resolve annihilation radiation from the very brightest subhaloes, which, typically, will contain stars.

\end{abstract}

\begin{keywords}
methods: numerical - Galaxy: structure - galaxies: spiral
\end{keywords}



\section{Introduction}

Dark matter accounts for more than $80\%$ of all matter in the Universe, but its nature is unknown. Historically, a particularly well motivated candidate for the dark matter particle has been a Weakly Interacting Massive Particle, perhaps the lightest supersymmetric partner of the known particles \citep[WIMPs, e.g.][]{BHS05}. Such particles may produce observable electromagnetic radiation through annihilation, which, for standard SUSY  WIMPS, produces $\gamma$-ray emission at energies of many GeV\citep[see e.g.][]{Arcadi+18}. The indirect detection of dark matter through this channel has been the subject of much research, but no clear signal, unambiguously due to dark matter, has so far been confirmed \citep{G16}.

The Large Area Telescope (LAT) onboard the Fermi Gamma-ray Observatory provides detailed observations of $\gamma$-ray emission across the entire sky \citep{Fermi09} and has been used to study possible dark matter annihilation signals from  dwarf satellites of the Milky Way \citep[e.g.][]{Albert+17} and from the diffuse component of its main dark matter halo \citep[e.g.][]{BBM94,BUB98,Ackermann+12,CLM18}. This latter component is brightest towards the Galactic Centre, but is severely contaminated by other sources of emission such as that from cosmic rays \citep{Ackermann+15} undergoing  hadronic interactions with other matter (leading, for example, to neutral pion decay). For this reason, observational analysis avoiding the Galactic Plane can have increased effective sensitivity, because the annihilation surface brightness remains large whereas contaminating emission is much reduced \citep{SWS03}. Nevertheless, detailed modelling of the complex and poorly known structure of the contamination is still critical for estimating or limiting the contribution of dark matter annihilation, so that foreground modelling remains the major source of uncertainty in the  results. 

A central prediction of the currently favoured $\Lambda$-Cold Dark Matter ($\Lambda$CDM) model of cosmology is the existence, within all galaxy haloes, of subhaloes with masses reaching down to Earth mass or smaller \citep[see][for a review]{FW12}. The most massive of these may contain satellite galaxies, but below some threshold essentially all subhaloes are expected to be dark. Such subhaloes are nevertheless predicted to be a significant source of annihilation emission, and it has been argued that the lowest mass objects may dominate the overall luminosity as a result of their large abundance and the high dark matter densities predicted for their inner regions \citep{CM00, BDV03, SWF08}. Numerical simulations provide the only feasible way to estimate accurately the abundance, internal structure and distribution of subhaloes within haloes like that of the Milky Way \citep{MGG99, KK99, GWJ04, DKM08} but for such haloes, dynamic range issues make it impossible to model subhaloes significantly below the threshold to host a satellite galaxy. Indeed, even for isolated haloes, techniques have only recently been developed which allow simulation of their present-day structure across the full mass range expected in the $\Lambda$CDM cosmology \citep{WBF20}. In addition, its dependence on the square of the local dark matter density makes annihilation luminosity very sensitive to the central structure of subhaloes, the region which is most difficult to simulate reliably. As a result, the predicted annihilation luminosity distribution in galaxy haloes, in particular the contribution from low-mass subhaloes, remains very uncertain. For, example \citet{SWF08} studied  substructure in a very high resolution, dark-matter-only simulation of a Milky Way mass halo, concluding that viewed from afar more than 99\% of the annihilation luminosity of the Milky Way would come from subhaloes too small to contain stars and situated far from the Galactic Centre. As seen from the Sun, they found that this component would still dominate the total annihilation flux, but would be distributed rather uniformly over the sky and so would be hard to distinguish from an extragalactic background. They argued that the most easily detectable component would likely be that from the halo's smooth dark matter distribution, although, as already noted, robust detection is hostage to reliable modelling of other sources of emission from the inner Galaxy.

None of the simulation work described so far considers the effect of baryons on dark matter haloes and subhaloes. In recent years, hydrodynamical simulations of galaxy formation have included processes such as gas cooling and condensation, star and supermassive black hole formation, feedback from supernovae and active galactic nuclei, and galactic winds, and have made significant progress towards producing realistic, disc-dominated, star-forming spiral galaxies like the Milky Way \citep[e.g.,][]{GC11,MPS14,Wetzel+Hopkins+Kim+16,BOM20}, as well as towards reproducing the galaxy population in representative  cosmological volumes \citep{SCB15,DPP16,PNH18}. Some of these simulations have been employed to estimate both direct \citep{BCS16,BFC19,BFF20} and indirect \citep[e.g.][]{SFT16,LBB17,LBB19} signals from dark matter. Of particular relevance for the work reported below, \citet{CBD20} studied a set of magneto-hydrodynamical simulations of the formation of individual Milky Way-like galaxies \citep[the \textlcsc{Auriga} simulations,][]{GGM17} and showed that the massive discs that grow within such systems cause a ``baryonic contraction'' of the central regions of their dark matter haloes as material is pulled inwards by the gravity of the discs \citep[see also][]{SFB15,LPG18}. This must increase the annihilation luminosity of the smooth dark matter relative to models that neglect baryonic effects. In turn, the increased halo concentration, together with the disc itself, enhances the tidal disruption of subhaloes that pass through the inner Galaxy, lowering their abundance and their annihilation/decay luminosity \citep[e.g.][]{SPJ17,GWB17,RFJ20}. This is a particularly strong effect for the subhaloes close to the Sun that have the highest predicted fluxes. Combined, these two effects cause flux predictions for dwarf galaxies and dark subhaloes to be much lower relative to the inner regions of the Milky Way's halo, once baryonic effects are properly accounted for. The present study aims to evaluate these effects quantitatively.

Here, we use the six highest resolution simulations in the \textlcsc{Auriga} suite to study how the annihilation fluxes predicted for Milky Way subhaloes relative to the main halo component are affected by baryonic processes. These magneto-hydrodynamic simulations resolve subhaloes down to $\sim 10^6 \rm M{\odot}$ and produce spiral  galaxies similar to the Milky Way. Each is available also in a dark-matter-only variant which has identical initial conditions and numerical parameters except for the exclusion of baryons. This allows us to show that baryonic processes increase the annihilation flux from the main halo component by about a factor of three, and reduce the typical flux from subhaloes by almost an order of magnitude. We also use recent results on the structure and annihilation properties of low-mass haloes to extrapolate our results all the way down to Earth-mass, finding upper limits on the {\it total} flux from subhaloes which are substantially below that of the main halo even seen from afar, and especially when seen from the Sun. This disagrees with earlier work which suggested the subhalo flux would dominate in both cases \citep[e.g.][]{SWF08}.

Our paper is organised as follows: we describe the simulations that we analyse in Section~\ref{sim} and the methodology used to calculate the annihilation luminosity of the main Milky Way halo and subhaloes in Section~\ref{method}. We present the results of our analysis in Section~\ref{results}, and summarise our findings in Section~\ref{conclusions}.

\section{Simulations}
\label{sim}

We here analyse a suite of simulations of six Milky Way-like systems and their local environments, taken from the \textlcsc{Auriga} project \citep{GGM17,GHF18}. These systems were specifically selected to have halo masses between $1$ and $2\times 10^{12} \rm M_{\odot}$,\footnote{We define halo mass, $M_{200}$, as the mass within the radius $R_{200}$ that encloses a mean  density 200 times the critical value for closure} and to be moderately isolated; they were identified in the $z=0$ snapshot of a dark-matter-only cosmological simulation of a periodic box of comoving size 100 Mpc, assuming a standard $\Lambda$CDM cosmology. The adopted cosmological parameters were $\Omega _m = 0.307$, $\Omega _b = 0.048$, $\Omega _{\Lambda} = 0.693$ and a Hubble constant of $H_0 = 100 h$ km s$^{-1}$ Mpc$^{-1}$, where $h = 0.6777$, taken from \citet{PC13}. At $z=127$, the resolution in dark matter particles of each halo and its surroundings is increased, and gas is added to create the initial conditions for a new ``zoom" simulation; this is evolved to the present day using the magnetohydrodynamics code \textlcsc{AREPO} \citep{Sp10}. 

Galaxy formation processes included in the ``full-physics'' versions of these simulations include self-gravity of all components, dissipative hydrodynamics, radiative cooling of gas, a two-phase model for cold and hot gas in star-forming regions, star formation, mass and metal return from stellar evolution, supermassive black hole formation, accretion and merging, energetic feedback from stars and AGN, and magnetic fields \citep[][]{GGM17}. The \textlcsc{AURIGA} model produces realistic spiral galaxies that are broadly consistent with a number of observations, in particular, with the star formation histories,  stellar masses, sizes and rotation curves of nearby galaxies like the Milky Way \citep{GGM17}. More detailed properties of such galaxies are also matched, for example, the distribution of HI gas \citep{MGP16}, their stellar halo properties \citep{MGG19}, the warps in their stellar discs \citep{GWG16}, the abundance and properties of the bars in their discs \citep{Fragkoudi+Grand+Pakmor+19}, the size and structure of their bulges \citep{GMG19}, the number and star formation histories (SFHs) of their dwarf satellite galaxies \citep{SGG17,DNF19}, and the properties of the magnetic fields in their discs \citep{PGG17,PGP18}. Individual simulations in the \textlcsc{Auriga} set match striking features specific to the  Milky Way, in particular: the Monoceros Ring \citep{GWM15}; the thin disc/thick disc dichotomy in element abundances \citep{GBG18}; the boxy-peanut bulge \citep{Fragkoudi+Grand+Pakmor+19}; and the radially anisotropic \emph{Gaia}-Enceladus or ``Sausage'' feature in the stellar halo \citep{FBD19,GKB20,BSF20}. Finally, large cosmological box simulations (that trade resolution for statistics) with a similar galaxy formation model reproduce many of the global properties of the observed galaxy population, such as the galaxy stellar mass function, galaxy sizes, the galaxy morphological mix, and the history of the cosmic star formation rate density \citep[e.g.][]{PNH18,NPS18}. 

The six systems we study here have resolution elements of
$\sim 6 \times 10^3$ $\rm M_{\odot}$ and $\sim 5 \times 10^4$ $\rm M_{\odot}$ for baryons and for dark matter, respectively, in the full-physics runs, and a dark matter particle mass of $\sim 6 \times 10^4$ $\rm M_{\odot}$  in the DMO runs. In both types of run, the gravitational softening length is $184$ pc after $z=1$ and is fixed in comoving units at earlier times. This resolution captures the formation of dark matter haloes and subhaloes of mass $\gtrsim 10^6$ $\rm M_{\odot}$ (corresponding to at least 20 dark matter particles per halo).

\section{Calculating annihilation luminosities}
\label{method}

\begin{figure*}
\includegraphics[width=\columnwidth,trim={0.2cm 0.2cm 0.5cm 0.5cm}, clip]{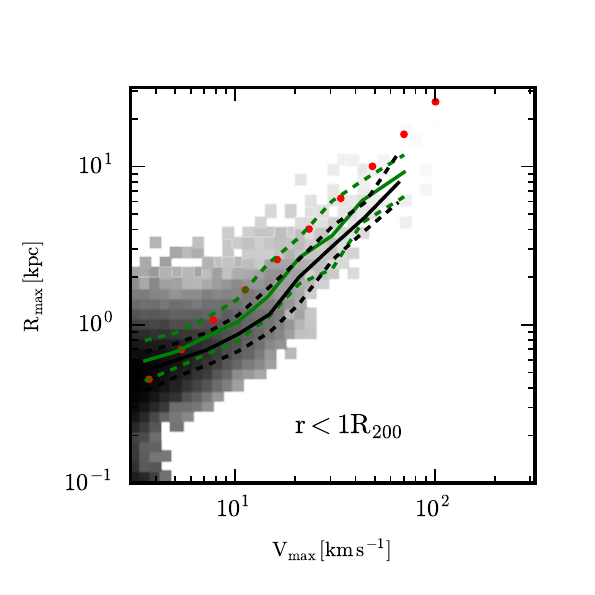}
\includegraphics[width=\columnwidth,trim={0.2cm 0.2cm 0.5cm 0.5cm}, clip]{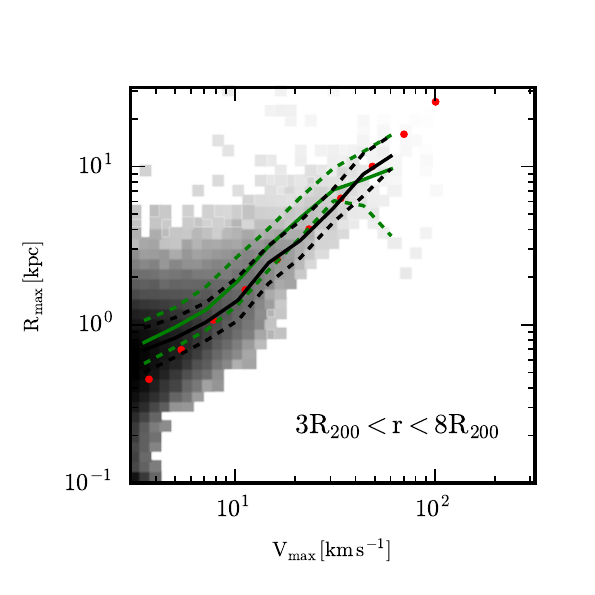}
\caption{The distribution in $R_{\rm max}$-$V_{\rm max}$ space of subhaloes within $R_{200}$ (left panel) and of haloes (i.e. centrals only) lying between 3 and 8 times $R_{200}$ from our main haloes (right panel). In each case, the 2d histogram in greyscale shows the distribution for our full-physics simulations, while the green lines indicate its 20\%, median and 80\% points for $R_{\rm max}$ at each $V_{\rm max}$. The black curves show the corresponding results for the DMO simulations. Red dots indicate the median relation found by \citet{WBF20} which we will use below to extrapolate our results to much lower subhalo mass.}
\label{vrmax}
\end{figure*}

\begin{figure*}
\includegraphics[width=\columnwidth,trim={0cm 0.2cm 0.5cm 0.5cm}, clip]{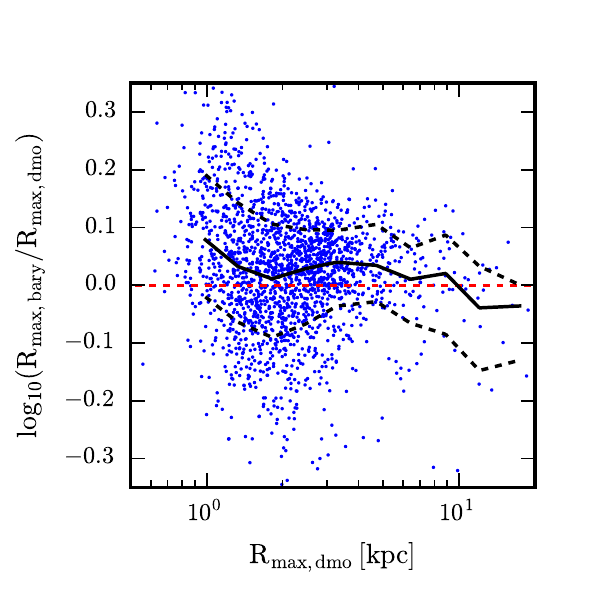}
\includegraphics[width=\columnwidth,trim={0cm 0.2cm 0.5cm 0.5cm}, clip]{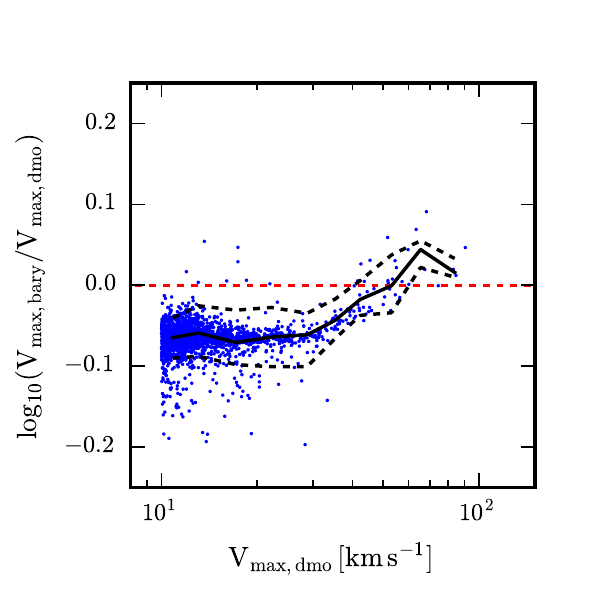}
\caption{Left: The ratio of $R_{\rm max}$ in the full-physics and DMO cases is shown as a function of its value in the DMO case for a sample of isolated haloes chosen as those in the right panel of Fig.\ref{vrmax} which are individually and unambiguously matched between the full-physics and DMO simulations and have $V_{\rm max} > 10$ $\rm km\,s^{-1}$ in the DMO simulations.  Right: For the same isolated haloes, the ratio of $V_{\rm max}$ in the two cases is shown as a function of its value in the DMO case. In both panels, the median relation is shown by a black solid curve and the 20\% and 80\% points by dashed curves. A dashed red line indicates the 1:1 relation.}
\label{matched}
\end{figure*}

\begin{figure}
\includegraphics[width=\columnwidth,trim={0.2cm 0.2cm 0.5cm 0.5cm}, clip]{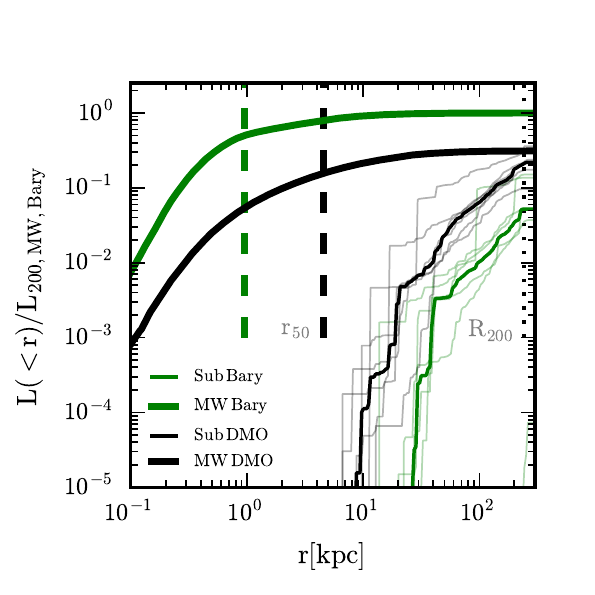}
\caption{The dark matter annihilation luminosity within radius $r$ as a function of that radius. Green curves give results from our six full-physics simulations while black curves give results for the corresponding DMO runs. The thickest curves show the mean luminosity for the smooth halo component (i.e. exclucing resolved subhaloes). The full-physics value at $R_{200}$ is used to normalise all luminosities in this figure. The vertical dotted line indicates $R_{200}$ which is almost identical in the two cases, while vertical dashed lines indicate the radii containing half of the total luminosity within $R_{200}$. The thin lines are cumulative luminosity curves for resolved subhaloes in individual simulations, while thicker lines show the medians of these curves over the six simulations in each set. }
\label{cumlum}
\end{figure}

This section describes how we estimate dark matter annihilation luminosities and fluxes for various components of our simulated galaxies. We begin by scaling the full-physics version of each simulation so that at $z=0$ it matches the Milky Way as well as possible. We define a factor $f$ which satisfies $f V_{\rm c,sim}(R_{\odot}/f) = V_{\rm c,\odot}$, where $V_{\rm c,sim}(R)$ is the circular velocity curve of the simulated galaxy in the plane of its disc, $R_{\odot} = 8.2$ kpc is the distance of the Sun from the Galactic Centre, and $V_{\rm c,\odot} = 240$ ${\rm km\, s^{-1}}$ is circular velocity of the Milky Way at $R_{\odot}$. All quantities in both the full-physics and the DMO simulation are then scaled using the same $f$: $M \to f^3M$; $\vec{x} \to f \vec{x}$; and $\vec{V} \to f \vec{V}$. For our six simulations $f$ varies between 0.93 and 1.10.

The luminosity density produced by dark matter annihilation at any point in space is given by $L=\mathcal{C}\rho ^2$, where $\mathcal{C}$ depends on particle physics (e.g., the velocity-weighted annihilation cross-section $\langle\sigma {\rm v}\rangle$ which we assume to be velocity independent, corresponding to an s-wave annihilation channel). For the purposes of this study, we express our results in terms of the relative luminosities and fluxes of different structures, and so, for convenience, we set $\mathcal{C}=1$. For the smooth component of the Milky Way's halo (i.e. after exclusion of all particles associated with identified subhaloes) we calculate the annihilation luminosity associated with the $i$th dark matter particle as the product of its mass, $m_i$, and $\rho _i$, the local density estimated from a Voronoi tessellation of the dark matter particle distribution. The luminosity of the Milky Way halo is then given by $L_{\rm MW} = \Sigma _i ^N \rho _i m_i$, where $N$ is the number of dark matter particles belonging to the smooth component of the Milky Way halo. 

Most of the annihilation signal for an individual dark matter halo or subhalo comes from far inside $R_{\rm max}$, the radius where $V_{\rm c}(R)$ reaches its maximum, $V_{\rm max}$. For the smaller resolved subhaloes in our simulations, these inner regions are not sufficiently well sampled for the Voronoi densities to be accurate, so that a procedure similar to the above typically substantially underestimates their total luminosity. To obtain a more reliable estimate, we therefore assume that their structure {\it interior} to  $R_{\rm max}$, follows a standard fitting formula. In this case, one can write,
\begin{equation}
L_{\rm sub} = C_{\rm Einasto} V^4_{\rm max} / G^2 R_{\rm max},
\label{eq1}
\end{equation}
where $G$ is the gravitational constant and $C_{\rm Einasto} = 1.87$ for an Einasto profile with $\alpha=0.16$. This profile is a good fit to the density profiles of isolated, present-day dark matter haloes of mass in the range $10^{-6}$ $\rm M_{\odot}$ to $ 10^{11}$ $\rm M_{\odot}$  \citep{WBF20}, so we are effectively assuming that when such haloes are accreted onto a larger object and become subhaloes, their inner structure is not substantially altered. 

For each subhalo, we calculate the circular velocity curve from the dark matter mass distribution over the range $r_{\rm soft} < r < 2 r_{\rm max}$ using equally spaced bins of 50 pc width, where $r_{\rm soft}$ is the softening length and $r_{\rm max}$ is the value returned by SUBFIND. We then carry out a least squares fit to an Einasto circular velocity curve (with the parameter $\alpha$ set to 0.16) to obtain values for $V_{\rm max}$ and $R_{\rm max}$. Below we also discuss isolated lower mass haloes outside of each of the main \textlcsc{AURIGA} objects. For these we obtain $V_{\rm max}$ and $R_{\rm max}$ values by applying the same procedure to their main subhalo as identified by SUBFIND.

\section{Results}
\label{results}

Fig.~\ref{vrmax} shows the $V_{\rm max}$ -- $R_{\rm max}$ relations both for subhaloes within $R_{200}$ and for isolated external haloes lying in the range $3 R_{200}< r < 8 R_{200}$, all of which are still within the high-resolution regions of the simulations. Results for the full-physics case are shown by the 2-D histograms, with green lines indicating the median, 20\% and 80\% values of $R_{\rm max}$ at given $V_{\rm max}$. Corresponding results for the DMO case are shown by the black lines.

The median relation for isolated haloes (right panel) in the DMO case can be compared to the ``high-mass'' end of the DMO relation of \citet{WBF20} (the red dots) which we will use later to extrapolate our results to masses far below our \textlcsc{AURIGA} resolution limit. Agreement is excellent down to $V_{\rm max}$ values below $10$ $\rm km \, s^{-1}$, at which point resolution  effects start to cause $R_{\rm max}$ to be overestimated in \textlcsc{AURIGA}. For  $V_{\rm max} \geq 10$ $\rm km \, s^{-1}$, the median relation for DMO subhaloes (left panel) is parallel to that for isolated haloes but the typical $R_{\rm max}$ is almost a factor of two smaller at given $V_{\rm max}$. This is a very similar result, although with a slightly larger offset, to that found by \citet{SWV08} for the DMO Aquarius simulations. The relations for the full-physics runs have very similar slope and scatter to the DMO relations but are shifted upwards by a factor of about 1.4 both for haloes and for subhaloes. A comparison of individual haloes in the two cases (see Fig.~\ref{matched}) shows that this reflects baryons failing to follow the distribution of dark matter in this mass range, resulting in slightly lower $V_{\rm max}$ and slightly larger $R_{\rm max}$ compared to the DMO relations. For $V_{\rm max} < 30~{\rm km\,s^{-1}}$ the relevant factors are about a 13\% reduction in $V_{\rm max}$ and an 8\% increase in $R_{\rm max}$ predicting, through equation~\ref{eq1}, a reduction in annihilation luminosity by a factor of 0.52.


Fig.~\ref{cumlum} shows separately the enclosed annihilation luminosity as a function of radius for the
smooth haloes (excluding resolved subhaloes) and for the resolved subhaloes. Results are plotted in green and black for the full-physics and DMO  simulations, respectively. The thick continuous curves are averages for the smooth component over the six simulations in each set. The thin stepped curves are for the subhalo components of individual simulations, while the thicker stepped curves show the medians for each set of six. All luminosity values are normalised to the median at $R_{200}$ for the smooth component of the full-physics runs. In the DMO runs, the luminosity from the smooth component rises steeply in the inner regions, and half of the total luminosity within $R_{200}$ is contained within $\sim 6$ kpc. The annihilation luminosity from resolved subhaloes rises much more slowly. The total within $R_{200}$ is typically comparable to that of the smooth component, but half of this value is reached only at $r\sim 100$kpc. These results agree quite well with those shown for a single, higher resolution DMO simulation by \citet{SWF08}, though their halo had a higher concentration than most of ours.

The addition of baryonic processes substantially alters these annihilation properties. The gravity of the stellar disc and bulge causes a significant contraction of the inner dark matter halo, driving up the luminosity of the smooth component by a factor of $\sim 3$ and reducing its half-light radius by a factor of $\sim 5$. The increased central concentration also increases the tidal disruption of resolved subhaloes with pericentres smaller than $\sim 20$ kpc, reducing the total luminosity from this component within $R_{200}$ by a factor of $\sim 6$
and its luminosity within smaller radii by larger factors. Combined, these two effects enhance the dominance of the smooth component by a factor of about 20 within $R_{200}$  and by even more at smaller radii. 

As noted in the introduction, previous work in this field has often argued that annihilation radiation from the Milky Way's halo is dominated by contributions from subhaloes very much less massive than those resolved by our simulations \citep{CM00,  BDV03, SWF08} and we wish to re-evaluate that suggestion here.
Although such \emph{subhalo} masses are far beyond the reach of current simulation techniques, recent advances have enabled the simulation of present-day {\it isolated} haloes over the full mass range of relevance \citep{WBF20}. The principal result of \citet{WBF20} is that halo structure is almost homologous at low mass; haloes in the range $10^{-5}< M_h/M_\odot<10^7$ all have similar density profiles with concentrations varying by less than a factor of 1.5. This implies that tidal stripping should also be approximately homologous when haloes in this mass range fall into the growing halo of the Milky Way, allowing us to use the scaling of halo properties at low-mass found by \citet{WBF20} to extrapolate our results for resolved subhaloes to much lower mass. Since annihilation radiation is predicted to come from far within $R_{\rm max}$, we use $V_{\rm max}$ rather than total mass as the variable characterising the overall scale of haloes and subhaloes.

As a first step we estimate the overall abundance of haloes across the full range expected for $V_{\rm max}$. We then renormalise this function to fit the abundance of low-mass but well resolved haloes and subhaloes in our simulations and use it to extrapolate their abundance to very small $V_{\rm max}$. Based on numerical data from the Millennium, Millennium-II and Millennium-XXL simulations, \citet{ASW12} suggested the following fitting function for halo abundance as a function of mass,
\begin{equation}
M \frac{dn}{dM} = \rho \frac{\rm d\,ln\,\sigma ^{-1}}{{\rm d}M} f(\sigma(M)),
\label{eq:massfn}
\end{equation}
where
\begin{equation}
f(\sigma (M)) = 0.201 \left[ \frac{2.08}{\sigma (M)} + 1 \right]^{1.7} \exp{\left[\frac{-1.172}{\sigma^2(M)}\right]},
\end{equation}
and $\sigma(M)$ is the rms linear fluctuation, extrapolated to $z=0$, within a spherical region which on average contains mass $M$.
The functional form here is based on theoretical work which suggests that it should extrapolate well to masses down to (but not including) the free-streaming cut-off (around $\sim 10^{-6}M_\odot$ for a 100 GeV WIMP). \citet{WBF20} found that below about $10^{11}M_\odot$ all isolated haloes have density profiles which are well described by the Einasto fitting-function with shape parameter $\alpha$ fixed to 0.16 and a typical concentration $c_{ein}=r_{-2}/R_{200}$ which they give as a function of $M_{200}$. This allows us to determine $V_{\rm max}$ as a function of $M_{200}$. Setting $M=M_{200}$ in equation \ref{eq:massfn},\footnote{\citet{ASW12} defined this function using Friends-of-Friends mass rather than $M_{200}$, but since in practice we will only use its shape at low mass where it is close to a power law, this is of no consequence for our analysis} we then obtain abundance as a function of $V_{\rm  max}$.

\begin{figure*}
\includegraphics[width=\columnwidth,trim={0.3cm 0.2cm 0.5cm 0.5cm}, clip]{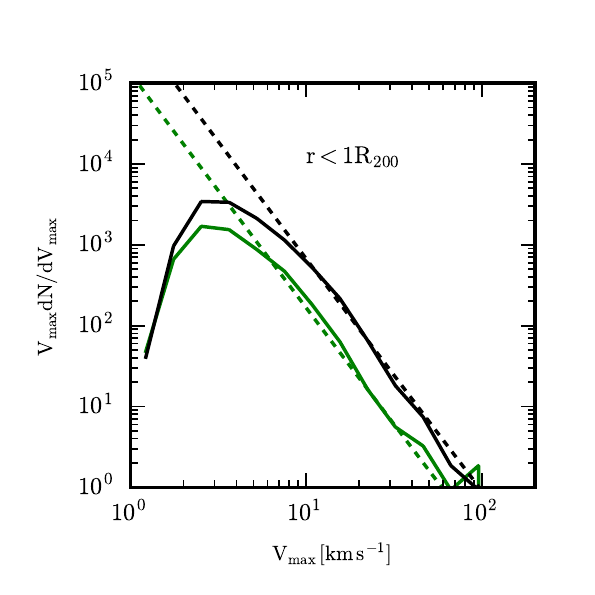}
\includegraphics[width=\columnwidth,trim={0.3cm 0.2cm 0.5cm 0.5cm}, clip]{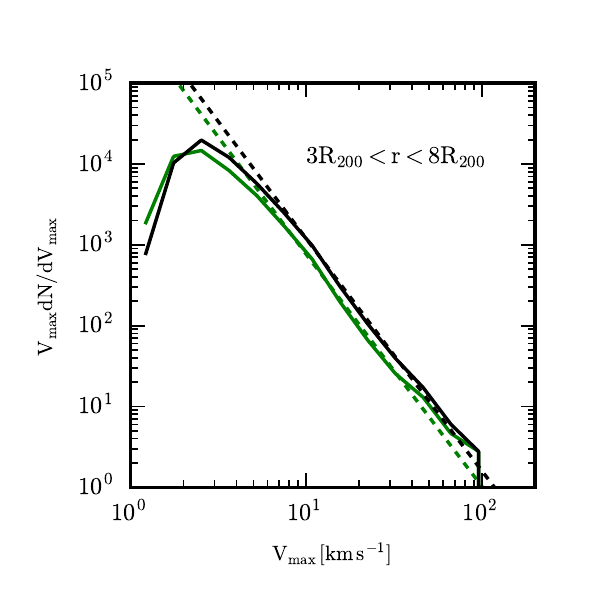}
\caption{Differential $V_{\rm max}$ functions for resolved subhaloes within $\rm R_{200}$ (left panel) and for resolved isolated haloes with $3\rm R_{200} < r < 8R_{200}$ (right panel) are shown by continuous green and black curves for the full-physics and DMO simulations, respectively. As described in the text, the dashed lines are obtained by combining the mass-concentration relation of \citet{WBF20} with the \citet{ASW12} differential mass function, and then renormalising to match the abundance of subhaloes/haloes in each resolved sample over the range $20 < V_{\rm max} / {\rm km\,s^{-1}} < 60$. This is then used to extrapolate our differential $V_{\rm max}$ functions below the resolution limit of the simulations. All curves give mean numbers per central Milky-Way-mass halo. }
\label{diffvmax}
\end{figure*}

In Fig.~\ref{diffvmax} we show the results in the form $V_{\rm max} dN/dV_{\rm max}$ both for subhaloes within $R_{200}$ (left panel) and for isolated haloes in the range $3<r/R_{200}<8$ (right panel). In each panel, solid curves show abundances measured directly from the simulations, black for the DMO case and green for the full-physics case. Dashed curves are the results obtained from equation \ref{eq:massfn} and the procedure just outlined, after renormalising to fit the solid curves
over the range $20~{\rm km\,s^{-1}} < V_{\rm max} < 60~{\rm km\,s^{-1}}$. Agreement is quite good above about $10~{\rm km\,s^{-1}}$, with some indication of small excesses in the simulations at large $V_{\rm max}$, particularly in the full-physics case where they may reflect shifts in the efficiency of rearrangement of the baryons. At  $V_{\rm max} < 10~{\rm km\,s^{-1}}$ the simulated curves drop below the extrapolations due to insufficient resolution. At given abundance, $V_{\rm max}$ for haloes is lower in the full-physics case than in the DMO case by a factor of 0.86 in good agreement with the shift seen for individual haloes in Fig.\ref{matched}. For subhaloes, the offset between the two cases is substantially larger, reflecting the enhanced tidal disruption of subhaloes in the full-physics simulations.

According to equation \ref{eq1}, estimating the annihilation luminosity of a small halo or subhalo requires a value for $R_{\rm max}$ as well as for $V_{\rm max}$. Fig.\ref{cumlum} included only resolved subhaloes and used the directly measured values of both parameters to estimate luminosities. However, Fig.\ref{diffvmax}  shows that the abundance is significantly underestimated for both haloes and subhaloes when  $V_{\rm max} < 10~{\rm km\,s^{-1}}$. Furthermore, $R_{\rm max}$ estimates become very noisy for such low-mass objects. In much of the following we therefore estimate the abundance of  these objects from extrapolations based on those in Fig.\ref{diffvmax}, and we estimate their $R_{\rm max}$ values, and hence their luminosities, as follows. Assuming an Einasto profile (with $\alpha=0.16$) the concentration-mass relation of \citet{WBF20} is easily converted into an $R_{\rm max}$ - $V_{\rm max}$ relation which Fig.\ref{vrmax} shows to fit very well the median relation for well resolved, isolated haloes in our DMO simulations. The median relations for other cases are similar but offset by multiplicative factors which we estimate from  Fig.\ref{vrmax} to reduce $R_{\rm max}$ at given $V_{\rm max}$ by 0.54  and  0.70 in the DMO subhalo case and the full-physics subhalo case, respectively, and to increase it by 1.40 for isolated haloes in the full-physics case. These relations then provide a median $R_{\rm max}$ for each $V_{\rm max}$ which we can insert into equation \ref{eq1} to get a median annihilation luminosity.

Armed with this relation between (median) luminosity and $V_{\rm max}$, it is  straightforward to convert the models for the differential $V_{\rm max}$ functions (the dashed lines in Fig.\ref{diffvmax}) into a differential luminosity functions $LdN/dL$. Finally, an Eddington-like correction needs to be applied to the results to account for the fact that there is significant scatter in $R_{\rm max}$  at given $V_{\rm max}$. Assuming this scatter to be lognormal with {\it rms} $\Delta\log_{10} R_{\rm max} = 0.17$ for all samples (based on Fig.\ref{vrmax}) and  approximating the luminosity functions as power laws with $LdN/dL\propto L^{-1}$, this correction shifts the luminosity functions towards higher luminosity by
$\Delta\log_{10} L = 0.03$.

\begin{figure*}
\includegraphics[scale=1.6,trim={0.3cm 0.cm 0.3cm 0.5cm}, clip]{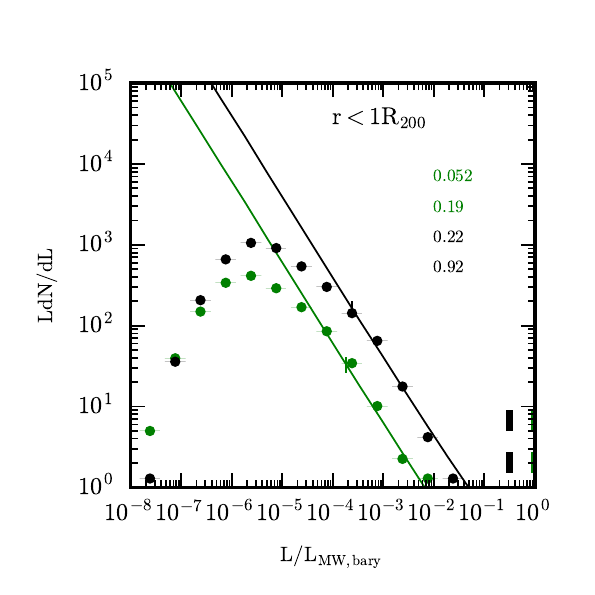}
\includegraphics[scale=1.6,trim={0.3cm 0.cm 0.3cm 0.5cm}, clip]{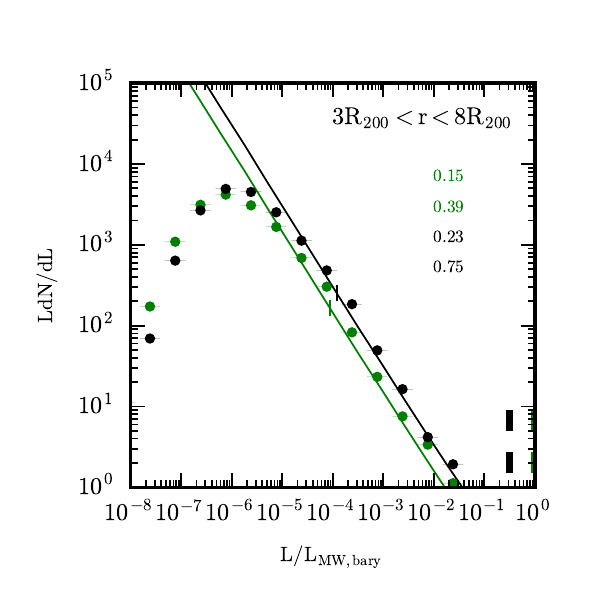}\\
\caption{Differential annihilation luminosity functions for resolved subhaloes within $\rm R_{200}$ (left panel) and for resolved isolated haloes with $3\rm R_{200} < r < 8R_{200}$ (right panel) are shown by green and black points for the full-physics and DMO simulations, respectively. The thin
lines are obtained as follows from the models for  $V_{\rm max} dN/dV_{\rm max}$ shown as dashed lines in
Fig.\ref{diffvmax}: the mass-concentration relation of \citet{WBF20} is converted into a $V_{\rm max}$ -- $R_{\rm max}$ relation; this is then offset by the amount needed to match the median $V_{\rm max}$ -- $R_{\rm max}$ relations for each sample in Fig.\ref{vrmax}; the result is used in equation \ref{eq1} to get luminosity as a function of  $V_{\rm max}$ which allows  $V_{\rm max} dN/dV_{\rm max}$ to be converted to $L dN/dL$; finally, an Eddington correction is made to account for the scatter in $R_{\rm max}$ at fixed $V_{\rm max}$. Vertical ticks on the lines show the point corresponding to $V_{\rm max} = 10~ {\rm km\,s^{-1}}$. The thick vertical dashed line at $L/L_{\rm MW,bary}\sim 0.3$ indicates the mean luminosity of the smooth halo component in the DMO runs. In each panel, pairs of colour-coded numbers indicate the total luminosity per
simulation (in units of $L_{\rm MW,bary}$) of each of the resolved (upper) and the full (lower) subhalo/halo  populations. All curves give mean numbers per central Milky-Way-mass halo.}
\label{lumfunc}
\end{figure*}

Fig.~\ref{lumfunc} illustrates the effects of these procedures. The filled circles show differential luminosity functions for simulated subhaloes (left panel) and for isolated haloes (right panel) calculated using the directly measured values of $V_{\rm max}$ and $R_{\rm max}$
as in Fig.\ref{diffvmax}. All annihilation luminosities in this figure are given, as before, in units of the mean value of the luminosity of the smooth component of the main halo in the full-physics runs. For comparison, the thin lines are calculated from the dashed lines in Fig.\ref{diffvmax} using the assumptions just outlined, namely that luminosity is a function of $V_{\rm max}$ according to equation \ref{eq1} with 
$R_{\rm max}$ a function of $V_{\rm max}$ as given by \citet{WBF20} offset as described in the last paragraph. A vertical tick on each of these lines indicates the point at which  $V_{\rm max}=10~{\rm km\,s^{-1}}$. Note that for the filled dots this correspondence is only approximate, since the simulations have scatter in $R_{\rm max}$ at each $V_{\rm max}$. This scatter is accounted for, as outlined above, in the model results shown by the thin lines. As expected, the agreement between the thin lines and the points for resolved halos/subhaloes is good at the bright end of all four luminosity functions.

The right panel of Fig.~\ref{lumfunc} shows that, for isolated haloes, the luminosity function for the full-physics
runs is offset towards lower luminosities by about the factor of 0.52 predicted from the results for individual matched haloes shown in Fig.\ref{matched}. For subhaloes, the offset is larger, about a factor of 5, as shown in the left panel of  Fig.~\ref{lumfunc}. This reflects the fact that these objects are subject not only to baryonic processes like cooling, star formation and feedback, which change their structure in much the same way as that of isolated haloes of similar mass, but also to enhanced tidal effects due to the increased central concentration of the mass distribution of the main halo. Notice that because this additional concentration {\it enhances} the luminosity of the smooth component of halo emission by a factor of about 3, the luminosities of the subhaloes {\it relative} to the smooth component are reduced by a factor of about 20 in the full-physics case.

The various luminosity functions of Fig.\ref{lumfunc} can be integrated to estimate the total annihilation luminosity from each of the four populations. For each case we give two numbers in the relevant panel, colour-coded to match the points and curves. The upper (smaller) number is that obtained by summing the luminosities of all resolved haloes/subhaloes (i.e. the total luminosity contributing to the plotted points) while the lower (larger) number is obtained by adding to the luminosity of all resolved  haloes/subhaloes with $V_{\rm max} > 10~ {\rm km\,s^{-1}}$ that found by integrating the corresponding thin line from the point where $V_{\rm max} = 10~ {\rm km\,s^{-1}}$ down to the free-streaming cut-off. The integration converges because of the rapid drop in concentration predicted by \cite{WBF20} as the cut-off is approached. In all cases, about 80\% of the total annihilation luminosity is predicted to come from objects below the resolution limit of the simulations. In the DMO case, the total subhalo luminosity within $R_{200}$ is 2.85 times that from the smooth component of the halo, whereas in the full physics case this ratio drops to just 0.19. These should be compared to the much larger ratio of 232 predicted by \citet{SWF08} which was a consequence of extrapolating with a model that predicts substantially higher concentrations for low-mass objects than was found by \cite{WBF20}.

\begin{figure*}
\includegraphics[scale=0.64,trim={1.1cm 5cm 1.1cm 5cm}, clip]{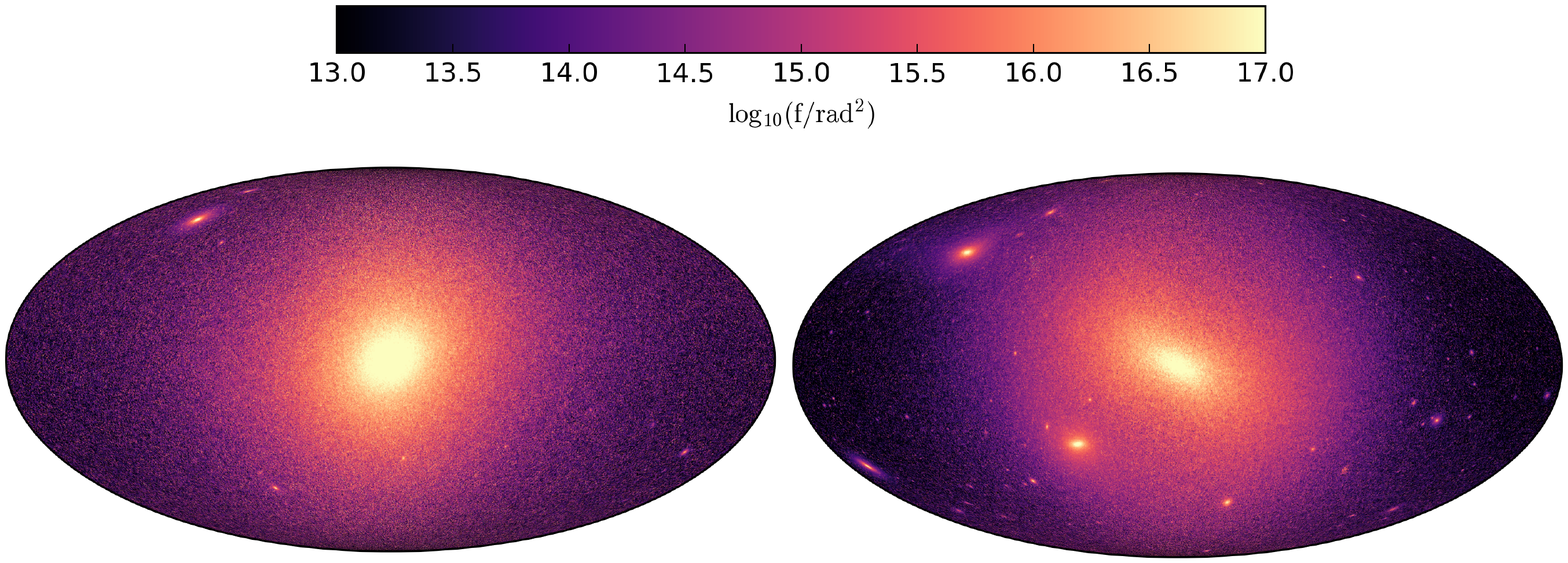}
\caption{All-sky maps of the flux density as seen from one Solar position in one full-physics simulation (left) and in its DMO counterpart (right); the dominant background annihilation signal originates from the smooth component of the halo. It is strongly concentrated towards the galactic centre and is brighter \citep[and rounder in shape, e.g.][]{PFG19} in the full-physics case. Subhalo fluxes are clearly systematically fainter in the full-physics case. }
\label{allsky}
\end{figure*}

This strong suppression of the subhalo signal relative to that from the inner main halo is even more marked if we consider the annihilation signal as seen by an observer at the Sun's position near halo centre, rather than as seen by a distant observer. This is because enhanced tidal disruption causes the typical heliocentric distance of satellites to be greater in the full-physics than in the DMO case.  To illustrate these effects, we estimate annihilation flux as $f = L/d^2$, where $L$ is the luminosity of a subhalo or main halo mass element and $d$ is its heliocentric distance from one of 18  ``Solar'' positions equally spaced around a ring of radius $8.2$ kpc in the mid-plane of the galactic disc. For each such position, we calculate the flux from every element of the smooth galactic halo (i.e. within $R_{200}$ and excluding resolved subhaloes) and we sum these to make a flux map of the full sky and to compute a total flux. We denote the average of these total fluxes over all positions and all 6 simulations as $f_{MW}$. For each position in each simulation we also calculate a flux for every subhalo/isolated halo as $V_{\rm max}^4/(R_{\rm max}d^2)$, together with an angular size computed as described below.  We use the directly simulated objects for $V_{\rm max} > 10~ {\rm km\,s^{-1}}$ and the Monte Carlo samples described in the next paragraph for smaller $V_{\rm max}$. This allows us to makes 18x6 pairs of all-sky maps of the annihilation flux. One of these pairs is shown in Fig.~\ref{allsky}.  It is clear that in the full-physics case the main halo emission is brighter, more concentrated and more symmetric about the centre; the subhaloes, on the other hand, are systematically fainter than in the DMO case and barely show up against the strong background provided by the smooth component.

\begin{figure}
\includegraphics[width=\columnwidth,trim={0 0 0.3cm 0.3cm}, clip]{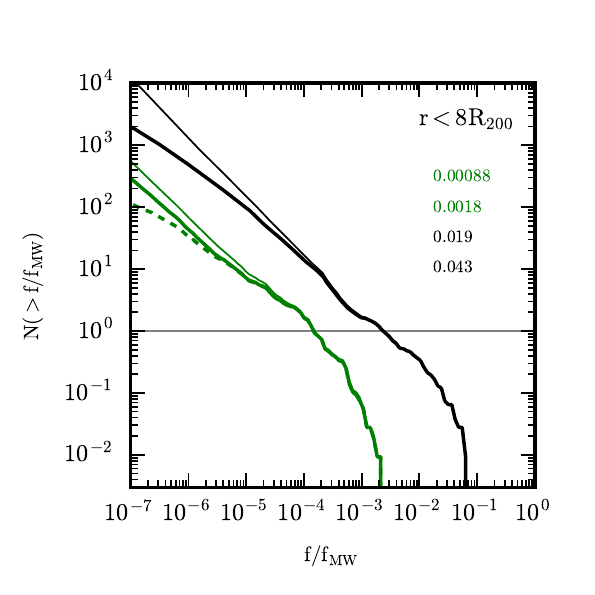}
\caption{The number of (sub)haloes per ``Milky Way'' analogue within 8$R_{200}$ and with annihilation flux $f$ exceeding a given fraction of $f_{MW}$,  the flux from the smooth component of the galaxy halo, as a function of that fraction. Each curve is compiled based on 18 ``solar'' viewing positions within each of six simulations. Results for our full-physics simulations are shown in green and for our DMO simulations in black. Thick curves are for (sub)haloes which are explicitly identified in the simulations, while thin curves use such (sub)haloes only when $V_{\rm max} > 10~ {\rm km\,s^{-1}}$ and extrapolate the properties of this well-resolved population to much lower $V_{\rm max}$ using Monte Carlo methods as described in the text. The four colour-coded numbers within the panel give the total flux (in units of $f_{MW}$) from resolved (sub)haloes (the upper, smaller numbers) and from all (sub)haloes down to the free-streaming limit ($\sim 10^{-6}M_\odot$; the lower, larger numbers). In the full-physics case, the dashed curve refers to subhaloes that contain at least some stars. For reference, $N(>f/f_{\rm MW})=1$ is marked with a grey horizontal line.}
\label{flux}
\end{figure}

\begin{figure}
\includegraphics[width=\columnwidth,trim={0.3cm 0 0.7cm 1.cm}, clip]{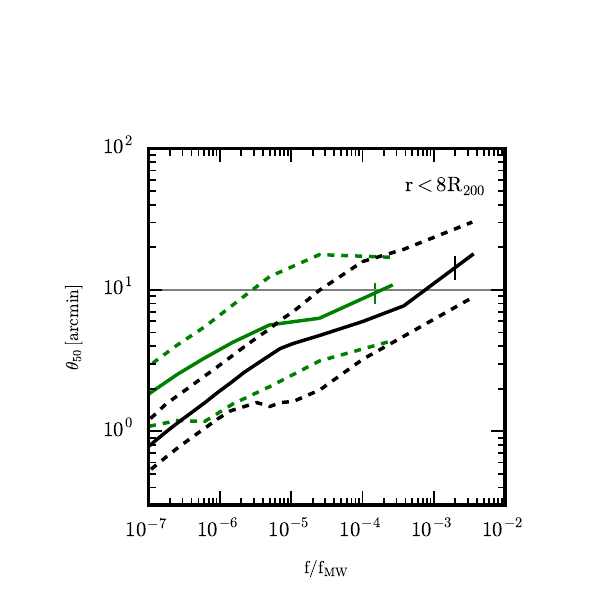}
\caption{Distribution of the angle on the sky subtended by the half-light radii of subhaloes as a function of their flux relative to that from the smooth halo component. As before black and green curves correspond to the DMO and full-physics simulations, and all (sub)haloes within $8R_{200}$ are considered when making the plot. Solid lines give the median value of $\theta_{50}$  at each value of $f/f_{MW}$, while dashed lines give its $\pm 1\sigma$ points. A vertical tick on each median line indicates the expected flux of the single brightest subhalo, while the horizontal line indicates the approximate FWHM of the Fermi-LAT for detecting objects of this type. }
\label{fsize}
\end{figure}

For making both these maps and the more quantitative measurements discussed  below, it is important that the distribution in galactocentric radius of objects too small to be resolved by our simulations should be as realistic as possible. We therefore extend the Monte Carlo sampling techniques described above as follows. For each of our ensembles of six simulations, we begin by making a stacked radial number density profile out to $8\rm R_{200}$ for resolved (sub)haloes with $10~{\rm km\,s^{-1}} < V_{\rm max} < 30~{\rm km\,s^{-1}}$, and we fit a smooth curve $n_{10,30}(r)$ to this profile.  For each of a series of thick radial shells of width $\Delta\log_{10} r\sim 0.2$ spanning $0.01< r/R_{200}<8$ we take the total number of resolved subhaloes in the chosen $ V_{\rm max}$ range and we use it to normalise a $V_{\rm max}$ distribution of the same shape as those used for extrapolation in Fig.\ref{diffvmax}. With these curves we then predict the expected number of objects as a function of $V_{\rm max}$ for $V_{\rm max} < 10~ {\rm km\,s^{-1}}$. These objects are distributed randomly in the angular coordinates and in radius according to a power-law fitted to the values of $n_{10,30}$ at the bin boundaries. Finally, each object is assigned an $R_{\rm max}$ value as described above when discussing Fig.\ref{lumfunc}.\footnote{We use the $V_{\rm max}$ -- $R_{\rm max}$ relations derived earlier for $3<r/R_{200}<8$ for all objects with $r>R_{200}$.} This produces a spherically symmetric Monte Carlo distribution of (sub)haloes with $V_{\rm max}$ distribution independent of radius and radial distribution paralleling $n_{10,30}(r)$ as measured from the simulations.

Subhalo populations predicted by this scheme for a single ``solar" position within a single simulation were used to make the maps in Fig.\ref{allsky}. However, since for each ensemble we have 18 ``solar'' positions for each of 6 simulations, our overall statistical power for inferring the expected distribution of subhalo fluxes is substantially better than this might suggest.  Thus in Fig.\ref{flux}, where we give our predictions for the cumulative distribution of satellite fluxes in units of $f_{MW}$, the mean flux of the smooth component, we can extend the numbers down to $N(>f/f_{MW}) = 0.01$. In this figure we show results both for resolved subhaloes alone, and for resolved subhaloes with $V_{\rm max} > 10~ {\rm km\,s^{-1}}$ together with MC subhaloes at lower $V_{\rm max}$. For the full-physics case, we also show a curve which includes only those subhaloes which contain stars. 

In Fig.\ref{flux}, unlike in Fig.\ref{lumfunc}, subhalo fluxes in each of our two ensembles (full-physics and DMO) are normalised by the flux of the smooth halo component {\it in that ensemble} rather than by the same quantity in both ensembles.  As a result, the difference between the full-physics and DMO cases appears larger here. Note also, that because a low-mass subhalo can have a large flux if it happens to be unusually close to the Sun, the extrapolated curves can differ from the full curves even at relatively bright fluxes. For the same reason, some relatively high-flux subhaloes are predicted to contain no stars.  The most likely flux of the (apparently) brightest subhalo (corresponding to $N(>f/f_{MW})=1$) is about $0.0023f_{MW}$ in the DMO case and only about $0.00015f_{MW}$ in the full-physics case. Finally, by extrapolating subhalo populations all the way down to $10^{-6}$ $\rm M_{\odot}$, we infer that the total flux from all small objects out to $8R_{200}$ is just 4.4\% and 0.18\% of that of the smooth component in the DMO and full physics cases, respectively; in each case, about half of this comes from subhaloes that are detected directly in the simulations. Again, this is much smaller than suggested in previous published work \citep[e.g.][]{SWF08} which assumed significantly higher concentrations for small (sub)haloes than was measured in the recent simulation of \citet{WBF20}. 

The detectability of annihilation radiation depends on the angular size of objects in addition to their flux. In order to compare predicted sizes to the resolution of current and future Fermi-LAT-type telescopes, we characterise the angular size of (sub)haloes on the sky by the angle containing half their light, $\theta _{50} = R_{50}/d$, where the projected half-light radius of an Einasto density profile can be found from the integral 

\begin{equation}
L(<R) = \int _0 ^R 2\pi R^\prime dR^\prime \int_0^\infty\big[\rho_{\rm Ein}({R^\prime}^2 + z^2)\big]^2 2dz, 
\label{eqL}
\end{equation}
where $\rho_{\rm Ein}$ is the Einasto profile. This gives: $R_{50} = 0.050 R_{\rm max}$ for $\alpha=0.16$. Fig.~\ref{fsize} shows the distribution of $\theta _{50}$ as a function of $f/f_{MW}$ over the flux range covered in Fig.~\ref{flux} for the samples indicated there by thin lines (i.e. resolved objects for $V_{\rm max} > 10~ {\rm km\,s^{-1}}$, MC objects at lower $V_{\rm max}$). We indicate the median and the $\pm 1\sigma$ points of the $\theta _{50}$ distribution at given flux by solid and dashed lines, respectively. Results are again colour-coded green and black for the full-physics and DMO runs respectively. A thin horizontal line indicates $\theta _{50} = 10$ arcmin, approximately the  FWHM of the pointspread function of the Fermi-LAT telescope for this kind of emission. Only the very brightest subhaloes are expected to be resolved by the Fermi-LAT.

The demographics of the highest flux subhaloes changes between the DMO and full-physics simulations: the median distance and mass of the brightest object
in our $18\times 6$ artificial skies is 20.1 kpc and $1.2\times10^{9}$ $\rm M_{\odot}$ in the DMO case, as compared with 33.3 kpc and $6.4\times10^{8}$ $\rm M_{\odot}$ in the full-physics case. This reflects the additional tidal disruption of objects passing through the inner galaxy in the full-physics case. The angular sizes are not so different in the two cases, however. We also note that the highest flux objects in the full-physics simulations usually, but not always, contain stars. Dark subhaloes can occasionally be close enough that their fluxes are comparable to those of the subhaloes containing dwarf galaxies. 

\section{Conclusions}
\label{conclusions}

Most published predictions for the distribution of annihilation radiation from dark matter in and around galaxies like our own have been based on high-resolution simulations of the formation of individual haloes that assume the standard $\Lambda$CDM paradigm but follow the dark matter component only. These have suggested: (i) that small subhaloes, for example those surrounding faint satellite galaxies, may be promising sources for a first detection in the Milky Way because of their high central densities and their lack of contaminating $\gamma$-ray sources; and (ii) that the total annihilation luminosity of the Milky Way may be dominated by that from very large numbers of subhaloes with individual masses many orders of magnitude below the resolution limit of the simulations.

In this paper, we re-evaluate these conclusions by analysing a suite of high-resolution $\Lambda$CDM simulations that produce galaxies similar to the Milky Way. Since each simulation is available in both a full-physics and a dark-matter-only (DMO) version, we are able to measure how structure in annihilation signal is affected by baryonic processes such as the formation of the Milky Way's stellar disc/bulge and ejection of gas from smaller satellite subhaloes. The effects are substantial; emission from the central regions increases and becomes more centrally concentrated, that from satellite objects  decreases by an even larger factor, especially as viewed from the Sun's position 8.2~kpc from the Galactic Centre.

The resolution of our \textlcsc{Auriga} simulations is still many orders of magnitude too poor to see the very low-mass subhaloes that have been suggested to dominate the total annihilation luminosity, but we can nevertheless address this issue by taking advantage of the recent breakthrough made by \citet{WBF20} who were able to simulate the $z=0$ structure of {\it isolated} $\Lambda$CDM haloes across the full mass range predicted for a standard SUSY WIMP ($10^{-6}M_\odot<M_h<10^{15}M_\odot$). They found that over most of this enormous range, present-day haloes have almost homologous density structure; their characteristic density varies very little, much less than assumed by most of the models used previously to extrapolate to low mass. Since homologous low-mass objects falling into a massive halo will be stripped homologously, independent of their actual mass, we can use the structural results of \citet{WBF20}
together with an isolated halo mass function \citep[we adopt that of][]{ASW12} to extrapolate our directly simulated results for subhaloes all the way down to the free-streaming limit, approximately Earth mass.

More specifically, our main conclusions are the following:
\begin{itemize}

\item{} In the \textlcsc{Auriga} simulations, the formation of the disc and bulge of the central galaxy adiabatically compresses the dark matter halo, enhancing its annihilation luminosity by a factor of three as seen both by a distant observer and from the Sun's position. The half-light radius of the emission is reduced by a factor of five as seen by a distant observer, and from $~\sim 29^\circ$ to $\sim 7^\circ$ as seen from the Sun.

\item{} As seen by a distant observer, the more massive satellite subhaloes are about six times less luminous in the full-physics simulations than in their DMO counterparts. Equivalently, satellites in the DMO simulations are a factor of six more numerous than those in the full-physics simulations above any given annihilation luminosity. This is partly due to baryon-induced changes in structure before satellite accretion, but mainly to enhanced tidal disruption of satellites by the greater central mass concentration in the full-physics case.

\item{} As seen from the Sun, the baryon-induced flux reduction is even larger, because satellites in the inner halo, relatively close to the Sun, are preferentially disrupted. This effect combines with the increased flux from the smooth component to substantially decrease the visibility of satellite subhaloes relative to the main halo. Thus, while in the DMO case the flux from the brightest subhalo is predicted to be $\sim 1/400$ that of the main halo, this ratio drops to $\sim 1/6000$ in the full-physics case. For the 10th or 100th brightest subhalo, the predicted flux ratios differ by a factor of 20 in the two cases.

\item{} From our extrapolation of the subhalo population all the way down to the free-streaming cut-off ($\sim 10^{-6}M_\odot$ for a SUSY WIMP of mass $\sim 100$~GeV), we find that, as seen by a distant observer, the total luminosity of all subhaloes within $R_{200}$ is only about 20\% of the smooth halo luminosity in the full-physics case, whereas it is almost three times larger than the smooth halo luminosity in the DMO case. In both cases, subhaloes directly resolved in our simulations contribute roughly a quarter of the total.

\item{} As seen from the Sun's position, the smooth halo is the dominant contributor to the annihilation flux in both cases. We find the total flux from all (sub)haloes within $8R_{200}$ to be just 0.2\% and 4\% of the flux from the inner halo in the full-physics and DMO cases, respectively, with roughly half of the emission coming from objects directly resolved in the simulations in each case. These numbers are much smaller than estimated by \citet{SWF08} because the subhalo concentrations we adopt for low-mass subhaloes (based on the simulation of \citet{WBF20})  are much smaller than those implied by the extrapolation underlying the earlier work.

\item{} All but the very brightest objects will be unresolved with the Fermi-LAT. These brightest objects are predicted to be $\sim 30$~kpc away and to have mass $\sim 6\times 10^8M_\odot$ in the full-physics case, somewhat more distant and lower mass than in the DMO simulations. Such an object might be expected to host an ultrafaint satellite galaxy.

\end{itemize}

In summary, our results suggest that the emphasis on the effects of substructure in much earlier work may have been misplaced. For example, we estimate that the total ``boost'' in annihilation luminosity for haloes of Milky-Way-like galaxies as a result of substructure on all scales is a factor 
of just 1.2. For the Milky Way itself, which we view from a position 8.2~kpc from the Galactic Centre, this boost factor is even smaller, just 1.002. These factors are negligible compared to other uncertainties in the problem. Similarly, the fact that the brightest subhalo is predicted to be about four orders of magnitude fainter than the inner Galaxy suggests that the latter will be detected robustly before the former becomes visible, even in the face of difficulties in removing contaminating sources of $\gamma$-radiation. 

Although our results should be quite robust for the massive subhaloes that are resolved in our simulations, there is still significant uncertainty in our extrapolation down to Earth mass. The simulation of \citet{WBF20} provides reliable data for {\it isolated} $z=0$ haloes over the full mass range, but we require additional assumptions to account for tidal stripping and disruption when such haloes become part of a larger system. An additional caveat comes from our assumption of an s-wave annihilation channel. Alternative particle physics models could give rise to a Sommerfeld enhancement, which would increase subhalo luminosities, relative to the main halo by one power of the ratio of $V_{\rm max}$ for the two systems, or even to p-wave or d-wave annihilation which would decrease subhalo luminosities relative to the main halo by two or four powers of the same ratio \citep[see][for example]{BKR19}. Finally, although we do not expect our conclusions to depend qualitatively on the baryonic physics model employed, other models may predict different degrees of halo contraction and subhalo destruction if the global properties (mass, size) and assembly history of the central galaxies are sufficiently different.  All these issues are clearly fertile ground for further work.

\section*{Acknowledgements}

We thank the referee for their prompt, constructive report.

\section*{Data Availability}
The data underlying this article will be shared on reasonable request to the corresponding author.

\bibliographystyle{mnras}
\bibliography{pap1.bib}

\bsp	
\label{lastpage}
\end{document}